\documentclass[a4paper,11pt]{article}
\usepackage{jinstpub} 
\usepackage{lineno}
\usepackage[version=3]{mhchem} 

\usepackage{comment}
\usepackage[version=3]{mhchem}
\usepackage[format=plain,labelsep=period]{caption}

\title{\boldmath Gas selection for Xe-based LCP-GEM detectors onboard the CubeSat X-ray observatory NinjaSat}

\author{T. Takeda$\rm^{1, 2}$, T. Tamagawa$\rm^{2, 3, 1}$, T. Enoto$\rm^{4, 2}$, T. Kitaguchi$\rm^2$, Y. Kato$\rm^2$, T. Mihara$\rm^{2, 3}$, \\W. Iwakiri$\rm^5$, M. Numazawa$\rm^6$, Y. Zhou$\rm^{1, 3}$, K. Uchiyama$\rm^{1, 3}$, Y. Yoshida$\rm^{1, 3}$, N. Ota$\rm^{1, 2}$, \\S. Hayashi$\rm^{1, 3}$, S. Watanabe$\rm^{1, 3}$, A. Jujo$\rm^{1, 3}$,  H. Sato$\rm^{7, 3}$, C.P. Hu$\rm^{8, 2}$, H. Takahashi$\rm^9$, H. Odaka$\rm^{10}$, T. Tamba$\rm^{11}$, K. Taniguchi$\rm^2$}

\affiliation{
$\rm^1$Tokyo University of Science, Department of Physics, 1-3 Kagurazaka, Shinjuku, Tokyo 162-8601, Japan \\
$\rm^2$RIKEN Cluster for Pioneering Research, 2-1 Hirosawa, Wako, Saitama 351-0198, Japan \\
$\rm^3$RIKEN Nishina Center, 2-1 Hirosawa, Wako, Saitama 351-0198, Japan \\
$\rm^4$Kyoto University, Department of Physics, Kitashirakawa Oiwake-cho, Kyoto, Japan\\
$\rm^5$Chiba University, International Center for Hadron Astrophysics, Chiba 263-8522, Japan\\
$\rm^6$Tokyo Metropolitan University, Department of Physics, 1-1 Minamiosawa, Hachioji, Tokyo 192-0397, Japan\\
$\rm^7$Shibaura Institute of Technology, Department of System Engineering and Science,307 Fukasaku, Minuma, Saitama, Saitama 337-8570, Japan\\
$\rm^8$National Changhua University of Education, Department of Physics, Changhua, 50007, Taiwan\\
$\rm^9$Hiroshima University, Department of Physics, 1-3-1 Kagamiyama Higashi-Hiroshima, Hiroshima 739-8526, Japan\\
$\rm^{10}$Osaka University, Department of Earth and Space Science, 1-1 Machikaneyama, Toyonaka, Osaka 560-0043, Japan\\
$\rm^{11}$The University of Tokyo, Department of Physics, 7-3-1 Hongo, Bunkyo, Tokyo 113-8654, Japan 
}
\emailAdd{tomoshi.takeda@a.riken.jp}

\abstract{
We present a gas selection for Xe-based gas electron multiplier (GEM) detectors, Gas Multiplier Counters (GMCs) onboard the CubeSat X-ray observatory NinjaSat.
To achieve an energy bandpass of 2--50 keV, we decided to use a Xe-based gas mixture at a pressure of 1.2 atm that is sensitive to high-energy X-rays. In addition, an effective gain of over 300 is required for a single GEM so that the 2~keV X-ray signal can be sufficiently larger than the electrical noise.
At first, we measured the effective gains of GEM in nine Xe-based gas mixtures (combinations of Xe, Ar, \ce{CO2}, \ce{CH4}, and dimethyl ether; DME) at 1.0~atm.  
The highest gains were obtained with \ce{Xe/Ar/DME} mixtures, while relatively lower gains were obtained with \ce{Xe/Ar/CO2}, \ce{Xe/Ar/CH4}, and Xe+quencher mixtures. 
Based on these results, we selected the Xe/Ar/DME (75\%/24\%/1\%) mixture at 1.2~atm as the sealed gas for GMC.
Then we investigated the dependence of an effective gain on the electric fields in the drift and induction gaps ranging from 100--650~V~cm$^{-1}$ and 500--5000~V~cm$^{-1}$, respectively, in the selected gas mixture.
The effective gain weakly depended on the drift field while it was almost linearly proportional to the induction field: 2.4 times higher at 5000~V~cm$^{-1}$ than at 1000~V~cm$^{-1}$. 
With the optimal induction and drift fields, the flight model GMC achieves an effective gain of 460 with an applied GEM voltage of 590~V.
}


\begin{document}
\maketitle
\flushbottom

\vspace{-8pt}
\section{Introduction}
\label{sec:intro}
\vspace{-1pt}
Gas electron multipliers (GEMs)~\cite{Sauli1997} are a type of micro-pattern gas detectors (MPGDs) that enable us to track charged particles with high spatial resolution.
Recently, GEM with liquid crystal polymer insulator (LCP-GEM)~\cite{Tamagawa2009} has been used in several cosmic X-ray polarimetry missions.
The main advantage of the LCP-GEM is its low outgassing, making it suitable for applications in satellite missions, where the sealed gas cannot be replaced for several years. LCP-GEM can also be used for non-imaging detectors because of its good gain uniformity across the GEM. 
In addition, compared to semiconductor detectors, GEM-based detectors can expand their effective area at a low cost and do not require cooling for operation. Therefore, non-imaging GEM-based detectors have the potential as detectors onboard CubeSat, where power, space, and budget are limited.

\begin{figure}[bht]
	\centering
	\includegraphics[width=140mm]{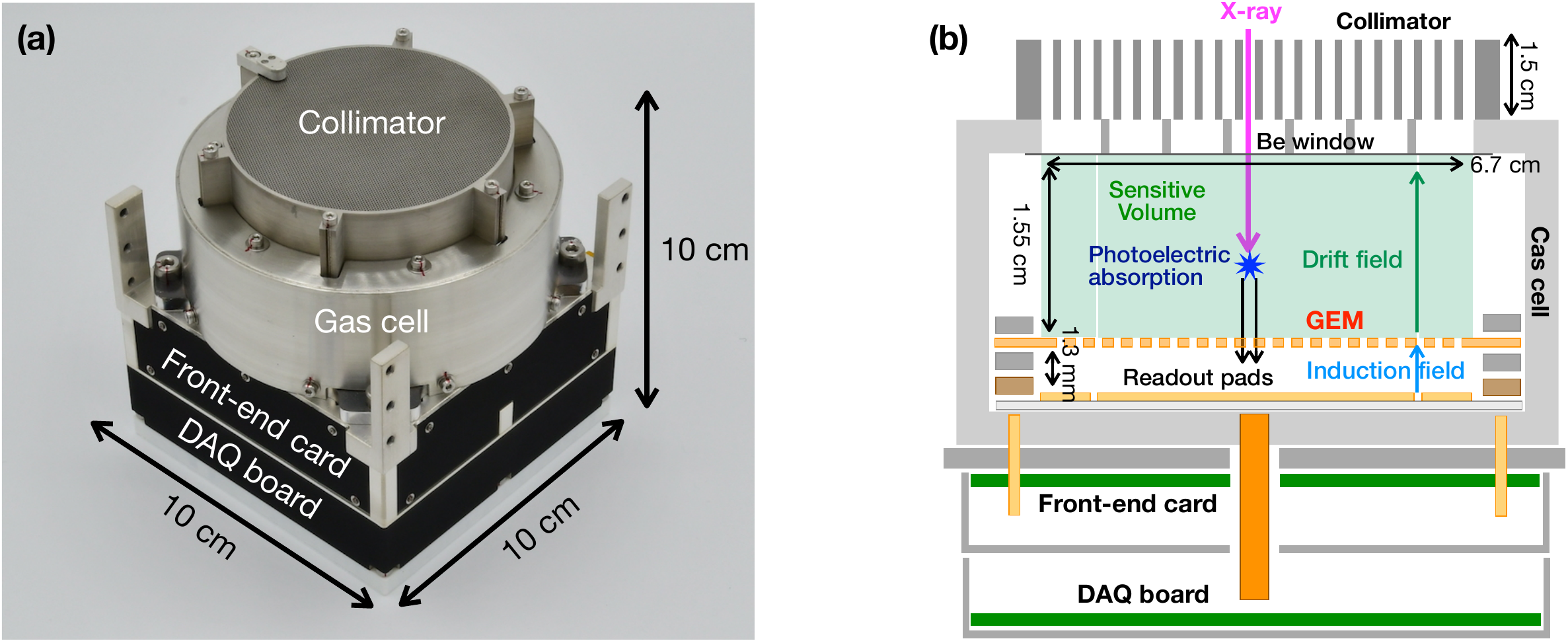}
	\caption{(a) Photograph of the flight model GMC. (b) Schematic cross-sectional view of GMC.}
    \label{fig:GMC}
\end{figure}

NinjaSat is a 6U-size (10$\times$20$\times$30~cm$^3$) CubeSat X-ray observatory, which will be launched in October 2023 and observes the time variability of bright X-ray sources in the 2--50~keV band~\cite{Enoto2020}. 
NinjaSat is equipped with two sets of Xe-based non-imaging gas X-ray detectors (Gas Multiplier Counters; GMCs).
Figure~\ref{fig:GMC} shows a photograph of the flight model GMC and its cross-sectional view.
To mitigate the risk of mission termination due to GEM failure caused by discharge, we adopt a single GEM configuration instead of the double or triple GEM configuration widely used in ground-based experiments.
When selecting the sealed gas for GMC, two essential things should be considered to realize the wide energy bandpass of 2--50~keV.
One is the use of a Xe-based gas mixture to be sensitive to high energy X-rays up to 50~keV. 
The other is the use of a gas that achieves a high effective gain of GEM so that 2~keV X-ray signals are sufficiently larger than the circuit noise.
In addition, because GMC does not have a pressure-supporting structure inside the gas cell, the sealed gas pressure must be higher than 1.2~atm so that the Be window would not be damaged due to differential pressure.
We set the pressure at 1.2~atm to achieve a higher gain at a low applied voltage to the GEM based on the anti-correlation between the effective gain of GEM and the gas pressure.
For the above reasons, we require an effective gain larger than 300 in a Xe-based gas mixture with a pressure of 1.2~atm at 0$^\circ$C.
However, in the case of a single GEM, it was known that the Xe-based gas mixture used in the past could not achieve sufficiently high gain to meet our requirement~\cite{Orthen2003}. 

This paper describes the GEM gain measurements in Xe-based gas mixtures performed to select an appropriate sealed gas for GMCs. In addition, only for the selected gas mixture of 75\% Xe, 24\% Ar, and 1\% dimethyl ether (DME) at 1.2~atm, the dependence of an effective gain on the electric fields in the drift gap, $E_{\rm d}$, and induction gaps, $E_{\rm i}$, are also described.

\section{Experimental setup}
We measured the effective gains of GEM in nine Xe-based gas mixtures (combinations of Xe, Ar, \ce{CO2}, \ce{CH4}, and dimethyl ether; DME) at 1.0~atm. For the GEM gain measurements, we used a gas cell with a maximum pressure resistance of 1.0 atm. 
Although this gas cell differs from the one used for the GMC, it has a GEM of the same design as GMC with a thickness of 100~$\rm \mu$m, a hole diameter of 70~$\rm \mu$m, and a pitch of 140~$\rm \mu$m. We utilized this gas cell to prevent excessive stress on the flight-equivalent GMC caused by repeated evacuations and refills with various candidate gases.
$E_{\rm d}$ and $E_{\rm i}$ were set to be 403 and $\rm 1488~V\ cm^{-1}$, respectively.  
The detector was irradiated with the X-rays from an X-ray generator with an electron-accelerating voltage of 40~kV and a target material of Mo.
The electron signal read out via the GEM cathode was fed into a charge-sensitive preamplifier (ORTEC 109) and a shaping amplifier (ORTEC 572). 
The signals were then digitized with a 12-bit ADC.
The effective gain of GEM is defined as the number of electrons read with the GEM cathode divided by the number of secondary electrons created by X-rays of Mo K$\alpha$.
The former electron number was calculated from the peak channel of the spectrum, while the latter was calculated assuming a mean energy for ion-electron creation of 22~eV, which is the value of the most abundant Xe in the gas mixtures.

The gain dependence on $E_{\rm d}$ and $E_{\rm i}$ were measured using a gas cell equivalent to the flight model GMC as shown in Figure~\ref{fig:GMC}. The gas cell was filled with a XeArDME (75\%/24\%/1\%) mixture at 1.2~atm, which was selected as the sealed gas for GMC based on the gain measurements. 
The gas cell was irradiated with the 5.9~keV X-rays from the radioactive \ce{^{55}Fe} source. 
We measured the charge collected with readout pads while scanning the drift and induction fields independently in the range of 100--650 and 500--5000~V~cm$^{-1}$, respectively.

\section{Results and Conclusion}
\begin{figure}[t]
	\centering
        \includegraphics[width=150mm]{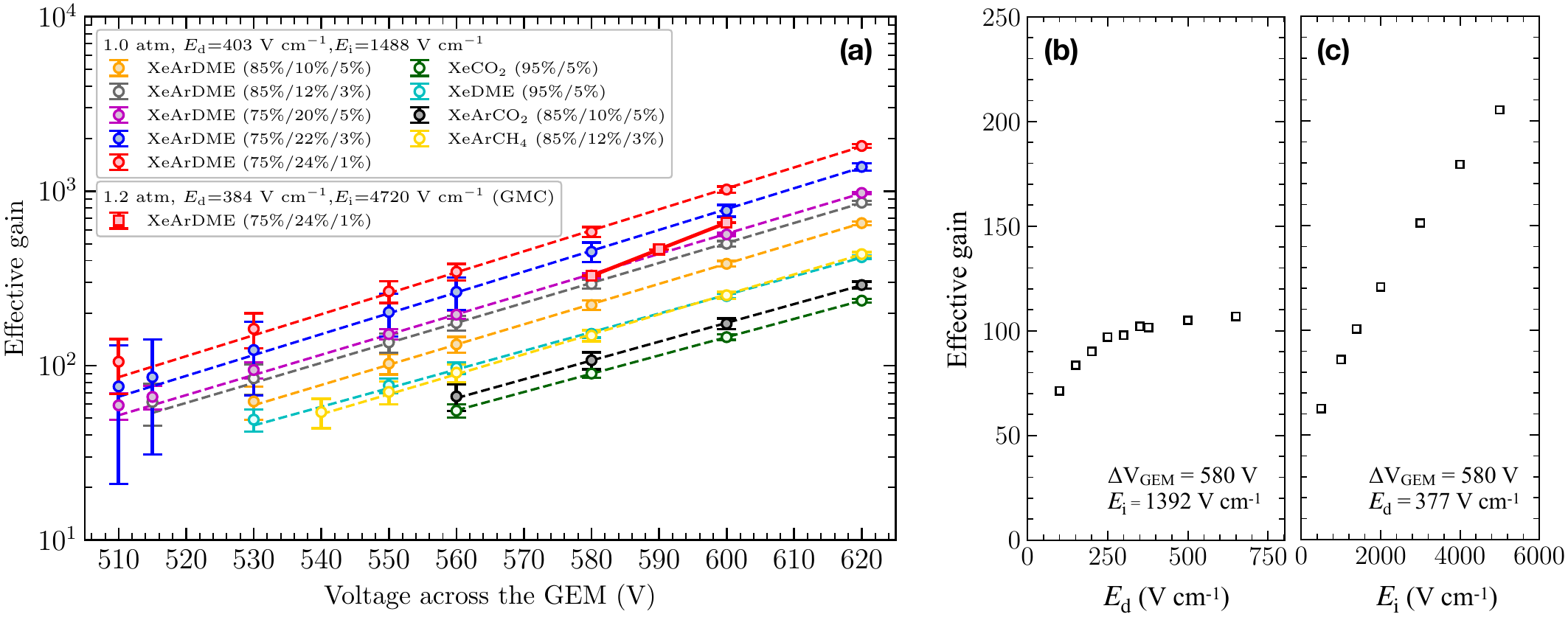}
	\caption{(a) Effective gain as a function of the applied GEM voltage in Xe-based gas mixtures. (b) Gain dependence on $E_{\rm d}$ and (c) $E_{\rm i}$ in the XeArDME (75\%/24\%/1\%) mixture at 1.2~atm.}
    \label{fig:GainCurve}
\end{figure}

Figure~\ref{fig:GainCurve}~(a) shows the effective gain of single LCP-GEM as a function of the applied voltage to the GEM ($\rm \Delta V_{GEM}$) in nine Xe-based gas mixtures at 1.0~atm.
Compared at the same voltage, the highest gain of $2\times 10^3$ was obtained with XeArDME (75\%/24\%/1\%) mixture, while relatively lower gains were obtained with \ce{Xe/Ar/CO2}, \ce{Xe/Ar/CH4}, and Xe+quencher mixtures. 
The highest gain of the XeArDME mixture is likely due to the penning effect between Ar, DME, and Xe.
Based on these results, we selected a XeArDME (75\%/24\%/1\%) mixture at 1.2~atm as a sealed gas for GMC.

The results of the $E_{\rm d}$ and $E_{\rm i}$ scan with the selected gas mixture are shown in Figure~\ref{fig:GainCurve}~(b) and (c), respectively. 
The effective gain was weakly dependent on $E_{\rm d}$ in the range of over 300~V~cm$^{-1}$, where the recombination of ion-electron pairs and electron attachment by electronegative atoms in the gas can be ignored.
On the other hand, the effective gain was almost linearly proportional to $E_{\rm i}$ because a larger $E_{\rm i}$ increases the charge collection efficiency. The effective gain was 2.4 times higher at $E_{\rm i}=5000$~V~cm$^{-1}$ than at 1000~V~cm$^{-1}$. 

Based on our study, the flight model GMC achieves an effective gain of 460 with a nominal operational settings of $\rm \Delta V_{GEM} = 590~V$, $E_{\rm d}=$ 384~V~cm$^{-1}$, and $E_{\rm i}=$ 4720~V~cm$^{-1}$ as shown in Figure~\ref{fig:GainCurve}~(a). 
In addition, GMCs achieve the energy bandpass of 2--50~keV with a total effective area of 32~cm$^2$ at 6~keV for two GMCs: two orders of magnitude larger than the previous X-ray detectors onboard CubeSats.

\appendix

\acknowledgments
This work was supported by RIKEN Junior Research Associate Program.





\end{document}